\newcommand{\be}{\begin{equation}}\newcommand{\ee}{\end{equation}}
\newcommand{\bea}{\begin{eqnarray}}\newcommand{\eea}{\end{eqnarray}}
\documentstyle[12pt]{article}
\def\PRL #1 #2 #3{{\sl Phys. Rev. Lett.} {\bf#1} (#2) #3}
\def\NPB #1 #2 #3{{\sl Nucl. Phys.} {\bf B#1} (#2) #3}
\def\NPBFS #1 #2 #3 #4{{\sl Nucl. Phys.} {\bf B#2} [FS#1] (#3) #4}
\def\CMP #1 #2 #3{{\sl Commun. Math. Phys.} {\bf #1} (#2) #3}
\def\PRD #1 #2 #3{{\sl Phys. Rev.} {\bf D#1} (#2) #3}
\def\PLA #1 #2 #3{{\sl Phys. Lett.} {\bf #1A} (#2) #3}
\def\PLB #1 #2 #3{{\sl Phys. Lett.} {\bf #1B} (#2) #3}
\def\JMP #1 #2 #3{{\sl J. Math. Phys.} {\bf #1} (#2) #3}
\def\PTP #1 #2 #3{{\sl Prog. Theor. Phys.} {\bf #1} (#2) #3}
\def\SPTP #1 #2 #3{{\sl Suppl. Prog. Theor. Phys.} {\bf #1} (#2) #3}
\def\AoP #1 #2 #3{{\sl Ann. of Phys.} {\bf #1} (#2) #3}
\def\PNAS #1 #2 #3{{\sl Proc. Natl. Acad. Sci. USA} {\bf #1} (#2) #3}
\def\RMP #1 #2 #3{{\sl Rev. Mod. Phys.} {\bf #1} (#2) #3}
\def\PR #1 #2 #3{{\sl Phys. Reports} {\bf #1} (#2) #3}
\def\AoM #1 #2 #3{{\sl Ann. of Math.} {\bf #1} (#2) #3}
\def\UMN #1 #2 #3{{\sl Usp. Mat. Nauk} {\bf #1} (#2) #3}
\def\FAP #1 #2 #3{{\sl Funkt. Anal. Prilozheniya} {\bf #1} (#2) #3}
\def\FAaIA #1 #2 #3{{\sl Functional Analysis and Its Application} {\bf
#1} (#2) #3}
\def\BAMS #1 #2 #3{{\sl Bull. Am. Math. Soc.} {\bf #1} (#2)
#3} \def\TAMS #1 #2 #3{{\sl Trans. Am. Math. Soc.} {\bf #1} (#2) #3}
\def\InvM #1 #2 #3{{\sl Invent. Math.} {\bf #1} (#2) #3}
\def\LMP #1 #2 #3{{\sl Letters in Math. Phys.} {\bf #1} (#2) #3}
\def\IJMPA #1 #2 #3{{\sl Int. J. Mod. Phys.} {\bf A#1} (#2) #3}
\def\AdM #1 #2 #3{{\sl Advances in Math.} {\bf #1} (#2) #3}
\def\RMaP #1 #2 #3{{\sl Reports on Math. Phys.} {\bf #1} (#2) #3}
\def\IJM #1 #2 #3{{\sl Ill. J. Math.} {\bf #1} (#2) #3}
\def\APP #1 #2 #3{{\sl Acta Phys. Polon.} {\bf #1} (#2) #3}
\def\TMP #1 #2 #3{{\sl Theor. Mat. Phys.} {\bf #1} (#2) #3}
\def\JPA #1 #2 #3{{\sl J. Physics} {\bf A#1} (#2) #3}
\def\JSM #1 #2 #3{{\sl J. Soviet Math.} {\bf #1} (#2) #3}
\def\MPLA #1 #2 #3{{\sl Mod. Phys. Lett.} {\bf A#1} (#2) #3}
\def\JETP #1 #2 #3{{\sl Sov. Phys. JETP} {\bf #1} (#2) #3}
\def\JETPL #1 #2 #3{{\sl  Sov. Phys. JETP Lett.} {\bf #1} (#2) #3}
\def\PHSA #1 #2 #3{{\sl Physica} {\bf A#1} (#2) #3}
\def\CQG #1 #2 #3{{\sl Class. Quantum Grav.} {\bf #1} (#2) #3}
\def\SJNP #1 #2 #3{{\sl Sov. J. Nucl. Phys. (Yadern.Fiz.)} {\bf #1} (#2) #3}
\def\a{\alpha}\def\b{\beta}\def\g{\gamma}\def\d{\delta}\def\e{\epsilon}

\def\l{\lambda}\def\L{\Lambda}
\def\om{\omega}
\newcommand{\p}[1]{(\ref{#1})}
\begin{document}
\renewcommand{\thefootnote}{\fnsymbol{footnote}}
\begin{flushright}
Preprint DFPD 95/TH/16 \\
hep-th/ \\
March, 1995
\end{flushright}

\vspace{1cm}
\begin{center}
{\Large \bf Note on  manifest Lorentz
and general coordinate invariance in duality symmetric
models\footnote{This work
was carried out as part of the
European Community Programme ``Gauge Theories, Applied Supersymmetry and
Quantum Gravity'' under contract SC1--CT92--D789, and
supported in part by M.P.I.}
}

\vspace{1cm}
{\bf Paolo Pasti},
\renewcommand{\thefootnote}{\dagger}
{\bf Dmitrij Sorokin\footnote{on leave from Kharkov Institute of
Physics and Technology, Kharkov, 310108, Ukraine.\\e--mail:
sorokin@pd.infn.it} and
\renewcommand{\thefootnote}{\ddagger}
Mario Tonin\footnote{e--mail: tonin@pd.infn.it}
}

\vspace{0.5cm}
{\it Universit\`a Degli Studi Di Padova
Dipartimento Di Fisica ``Galileo Galilei''\\
ed INFN, Sezione Di Padova
Via F. Marzolo, 8, 35131 Padova, Italia}

\vspace{1.5cm}
%\vspace{0.3cm}
{\bf Abstract}
\end{center}

\bigskip
We consider a  generalization of a duality
symmetric model proposed by Schwarz and Sen \cite{ss}. It is based on
enlarging the model with a dynamical vector field being a
time-like component of a local Lorentz frame. This allows one to
preserve the manifest Lorentz invariance of the model in flat
space--time. The presence of this field is regarded as a relic of
gravitational interaction which respects the general coordinate invariance
in curved space--time but breaks the local Lorentz symmetry in tangent
space down to its spacial subgroup.


\newpage
\renewcommand{\thefootnote}{\arabic{footnote}}

Following the course of studying duality in string theory and
related topics
Schwarz and Sen \cite{ss} proposed a class of gauge invariant
actions which are manifestly invariant under duality transformations.
``The price for doing this is the sacrifice of manifest Lorentz
invariance or general coordinate invariance, though both symmetries can
be realized nonetheless''. But as it has already happened not one time,
models, which at first have been considered as  manifestly Lorentz
non--invariant later on were reformulated in a Lorentz invariant way.
The Green--Schwarz superstring is one of these examples \cite{gs}.

In a recent paper \cite{l} Khoudeir and
Pantoja proposed a covariant form of the Schwarz--Sen action by
introducing into the theory
an auxiliary time--like constant vector. However, being a
constant, this vector, in fact, violates Lorentz invariance. Hence, in
spite of the nice covariant form of the action and equations of motion
Lorentz invariance have not been restored.

Below we propose a generalization of the duality symmetric action
\cite{ss,l}
by considering the auxiliary vector as a Lorentz frame vector
field whose presence in the model can be regarded as a relic of a
gravitational interaction. At the classical level this vector field
completely decouples from the gauge field sector.
We will see that the coupling of duality
symmetric fields to gravity respects the general coordinate invariance
in curved space--time but breaks the local Lorentz symmetry in tangent
space down to its spacial subgroup.

For simplicity we will deal with the dual symmetric version of
(supersymmetric)
Maxwell theory, though the construction is generalized to the case of
antisymmetric gauge fields  \cite{ss} as well.

The duality symmetric form of the D=4 Maxwell action proposed by
Schwarz and Sen involves two abelian gauge fields $A^\a_m$ ($\a$=1,2;
m=0,1,2,3)
which are present in the action on an equal footing:
\begin{equation}\label{ss}
S=-{1\over 2}\int d^4x(B^{i\a}{\cal L}^{\a\b}E^\b_i+B^{i\a}B_{i}^\a),
\end{equation}
where
\begin{equation}\label{eb}
E^\a_i=F^\a_{0i}=\partial_0A^\a_i-\partial_iA^\a_0,
\qquad
B^{i\a}={1\over 2}\varepsilon^{ijk}F^\a_{jk}=
\varepsilon^{ijk}\partial_jA^\a_k,
\end{equation}
$i,j,k = 1,2,3$ are spacial indices,
and ${\cal L}^{\a\b}$ is the antisymmetric
unit matrix $({\cal
L}^{12}=1)$.

Apart from the ordinary abelian gauge transformations of
$A^\a_m$ the action is invariant under
local transformations
\begin{equation}\label{o}
A^\a_0~\rightarrow~A^\a_0+\Psi^\a(x)
\end{equation}
and global
$SO(2)$ transformations which mix $A^1_m$ and
$A^2_m$, its discrete subgroup  being the duality
symmetry:
\begin{equation}\label{d}
A^\a_m~\rightarrow~{\cal L}^{\a\b}A^\b_m.
\end{equation}

Using the local symmetries and the $A^\a_m$ equations of motion one can
eliminate one of the gauge fields and get the conventional Maxwell
theory for the other one. Then the duality symmetry is reduced to the
duality between the electric and magnetic strength of the remaining
Maxwell field \cite{ss}.

The action is also invariant under the spacial rotations,
and has a global symmetry which is reduced to the Lorentz
transformations when the model is reduced to Maxwell theory \cite{ss}.

In the paper \cite{l} by use of a constant time--like vector $u^m$
subject to the condition
\begin{equation}\label{u}
u^mu_m=-1
\end{equation}
the action \p{ss} was rewritten in the following form
\begin{equation}\label{l}
S=-{1\over 2}\int d^4xu^mF^{*\a}_{mn}({\cal
L}^{\a\b}F^{\b np}-F^{*\a np})u_p,
\end{equation}
where $F^{*\a}_{mn}={1\over 2}\e_{mnpq}F^{\a pq}$ is the dual of
$F^{\a}_{mn}=\partial_mA^{\a}_n-\partial_nA^{\a}_m$.

In \p{l} the transformations \p{o} take the form
\begin{equation}\label{ou}
A^\a_m~\rightarrow~A^\a_m+u_m\Psi^\a(x).
\end{equation}

Since $u_m$ is a constant which satisfies \p{u} the action \p{l}
corresponds to a different choice of the Lorentz frame in space--time
and is reduced to \p{ss} by an appropriate Lorentz transformation
of $u_m$ to $u_m=-\d^0_m$. Hence, in spite of the Lorentz covariant form,
eq. \p{l} does not possess  manifest Lorentz symmetry.

To get a Lorentz
symmetric action one must consider $u_m$ as an $x$--dependent vector field
and take into account its equations of motion in the presence of the
orthonormality condition \p{u}. Then $u_m$ becomes a component of a
local frame $(u_m,~u^i_m)$ in the flat space--time, or a harmonic
\cite{gikos} of the Lorentz group,
where $(u_m,~u^i_m)$ satisfy
the orthonormality condition \p{u}
and
\begin{equation}\label{or}
u_mu^{im}=0,\qquad u^i_mu^{jm}=\d^{ij}.
\end{equation}
Implicit use of \p{or} below will substantially simplify finding the
symmetries and solving for the equations of motion of the model.
\footnote{Different types of the harmonic fields have been
successfully used to
solve covariance problems in various theories including superstrings.
Due to the great amount of literature on this subject we will
abstain from giving references.} In conclusion we will make a conceptual
comment on
the possibility of treating of $(u_m,~u^i_m)$ as a relic of
a gravitational field vierbein.

If $u_m$ is a field, the action \p{l} looses the local invariance
\p{ou} (or \p{o}), while the latter plays an important
role in establishing the
relationship of the model with Maxwell theory. Thus, eq. \p{l} must be
modified in a way which restores this symmetry.

We write down the generalized action in the following form:
\begin{equation}\label{gen}
S=\int d^4x \left(-{1\over 8}F^\a_{mn}F^{\a mn}+
{1\over 4}u^m{\cal F}^\a_{mn}{\cal F}^{\a np}u_p
-\e^{mnpq}\L_{mn}\partial_pu_q +\l(x)(u_mu^m+1)\right).
\end{equation}
 For constant $u_m$ eq. \p{gen} reduces to \p{l}. From the form
\p{gen} of the action one may see
where the duality symmetric model differs from
the ordinary theory of two independent Maxwell fields. The first term in
\p{gen} is just the sum of the two Maxwell terms. The nontrivial
``interaction'' between the two gauge fields is described by the second
term, where
\begin{equation}\label{sd}
{\cal F}^\a_{mn}={\cal L}^{\a\b}F^\b_{mn}-F^{*\a}_{mn},
\qquad
{\cal F}^\a_{mn}\equiv{1\over 2}\e_{mnpq}{\cal L}^{\a\b}{\cal F}^{\b pq}
\end{equation}
is the self--dual tensor. Note that because of this self--duality
${\cal F}^\a_{mn}{\cal F}^{\a nm}=0$ identically,
which causes the
problem of Lorentz covariant incorporation of \p{sd} into the action.

The third term  (where $\L_{mn}(x)$ is an auxiliary antisymmetric gauge
field) ensures the invariance of the action with respect to \p{ou}
provided $\L_{mn}$ transforms as follows:
\begin{equation}\label{la}
\L_{mn}~~\rightarrow~~\L_{mn}+\Psi^\a{\cal F}^\a_{mp}u^pu_n-
\Psi^\a{\cal F}^\a_{np}u^pu_m.
\end{equation}
One more, obvious, local symmetry  of the action is its invariance (up
to a total derivative)
under the following transformations of $\L_{mn}$:
\begin{equation}\label{dop}
\d\L_{mn}=\partial_m\L_n(x)-\partial_n\L_m(x).
\end{equation}

The fourth term, where $\l(x)$ is a Lagrange multiplier, ensures \p{u}.

The equations of motion obtained from the action \p{gen}
have the following form:
\begin{equation}\label{a}
\d A^\a_l:~~~~~
\epsilon^{lmnp}\partial_m(u_n{\cal F}^\a_{pr}u^r)=0,
\end{equation}
\begin{equation}\label{L}
\d\L_{lm}:~~~~~
\epsilon^{lmnp}\partial_nu_p=0,
\end{equation}
\begin{equation}\label{ue}
\d u_m:~~~~~
-{1\over 2}{\cal F}^{\a m}_{n}{\cal F}^{\a np}u_p+
\e^{mnlp}\partial_n\L_{lp}=
\l(x)u^m.
\end{equation}

In view of \p{L} the general solution to \p{a} is
\begin{equation}\label{sol}
{\cal
F}^\a_{pr}u^r=\partial_p{\tilde\Psi}^\a+u^r\partial_r(u_p\tilde\Psi^\a),
\end{equation}
where ${\tilde\Psi}^\a$ are arbitrary functions. But (again due to
\p{L}) the transformation of ${\cal F}^\a_{pr}u^r$ with respect to
\p{ou} has the same form as \p{sol}. So, by putting
${\Psi}^\a={\cal L}^{\a\b}{\tilde\Psi}^\b$
we can choose the gauge of \p{ou} in which ${\cal
F}^\a_{pr}u^r=0$ and, hence, ${\cal F}^\a_{pr}=0$ since it is self--dual.
As a result we arrive at the duality condition obtained in \cite{ss,l}:
\begin{equation}\label{dual}
{\cal L}^{\a\b}F^{\b}_{mn}={1\over 2}\e_{mnpq}F^{\a pq}.
\end{equation}
When \p{dual} is satisfied, the first term in the l.h.s.
of \p{ue} is zero.
Thus, on the mass shell the gauge field sector completely decouples from
$u_m$ and $\L_{mn}$,
and, in compliance with \cite{ss}, the former is reduced to
ordinary Maxwell theory.

{}From \p{L} we get
\begin{equation}\label{phi}
u_m=\partial_m\phi(x)
\end{equation}
(where \p{phi} must satisfy \p{u}),
while eq. \p{ue} reads that (in the gauge \p{dual})
the only non--zero component of the
strength vector $\e^{mnlp}\partial_n\L_{lp}$ of the gauge field
$\L_{lp}$ is parallel to $u_m$ with the proportionality coefficient to be
$\l(x)$:
\begin{equation}\label{p}
\e^{mnlp}\partial_n\L_{lp}=\l(x)u^m.
\end{equation}
 The  components of $\L_{lp}$ which do not contribute to \p{p} are
gauged away by the transformations \p{dop}. Hence, the only independent
field in \p{p} is $\l(x)$. The latter satisfies the
condition
\begin{equation}\label{con}
\partial_m(u^m\l)=0,
\end{equation}
which follows from \p{p}.

It is possible to show that, as in refs. \cite{ss,l}, one can eliminate
one of the gauge fields (for example $A^2_m$)
from the action \p{gen} using its equations of
motion and reduce \p{gen} to the ordinary Maxwell action plus  terms
which contain decoupled auxiliary fields:
\begin{equation}\label{max}
S=\int d^4x \left(-{1\over 4}F^1_{mn}F^{1 mn}
-\e^{mnpq}\L_{mn}\partial_pu_q +\l(x)(u_mu^m+1)\right).
\end{equation}

Thus, we have constructed the manifest
Lorentz invariant version of
a duality symmetric gauge theory which at the classical level contains
in addition to the Maxwell field
two decoupled redundant fields $\phi(x)$ \p{phi} and $\l(x)$ \p{p}.

Perhaps, one would desire to get rid of this fields, at least at the
classical level.
A possible way
to do this would be to find in the model, or in some
its generalization,
a hidden symmetry  which would allow one to gauge fix $u_m(x)$ to a
constant vector. The following speculations demonstrate that
there is indeed a hint of such a symmetry. The actions \p{gen}, \p{max}
are inert under infinitesimal transformations
\begin{equation}\label{sy}
\d u_m =\partial_m\varphi(x)
\end{equation}
$$
\d A^\a_m=\varphi{\cal L}^{\a\b}{\cal F}^\b_{mn}u^n,\qquad
\d\L_{mn}=\varphi{\cal F}^{\a r}_{m}
u_r{\cal F}^{\b s}_nu_s{\cal L}^{\a\b}
$$
provided
\begin{equation}\label{var}
u^m\partial_m\varphi=0.
\end{equation}
Eqs. \p{sy} are a localization of the global transformations of
the Schwarz--Sen action \cite{ss}
which substitute off--shell Lorentz invariance in their model (where
$\varphi=x^mv_m$ with $v_m$ being a constant velocity vector).

The weak point of the transformations \p{sy} is that the
infinitesimal parameter $\varphi(x)$ must be restricted by the condition
\p{var}
which is the consequence of \p{u}. As a result of
this restriction the r.h.s. of equation \p{ue} is not invariant under
\p{sy}, while the l.h.s. is invariant. Thus, in general, the
equations of motion of the model are not invariant under \p{sy}.
But if the classical value of $\l(x)$ is zero, then the transformations
\p{sy} do become a symmetry of the equations of motion \p{u},
\p{a}--\p{ue} and allow one to
gauge fix $u_m(x)=\partial_m\phi(x)$ (eq.\p{phi})
to be a constant vector with the unit norm \p{u}. At this ``critical"
point the present model completely coincides with the model of Schwarz
and Sen.

But more interesting is to learn
what kind of more general (supersymmetric) theory the
fields $u_m(x)$ and $\L_{mn}(x)$ came from. The natural suggestion is
to regard $\L_{mn}(x)$ as an axion potential and
$u_m(x)$ as a relic of the time--like component of a
gravitational field vierbein in a model of duality symmetric gauge
fields
interacting with gravity in such a way that the second and the third
term of \p{gen}
break the manifest local Lorentz invariance of the {\it tangent} space
down to $SO(3)$, while the general coordinate invariance of the {\it
curved}
space--time remaines intact. The simple calculus of independent
components of the vierbein testifies to this assumption. Indeed, the
conventional theory of General Relativity can be formulated equally well
in terms of a metric $g_{mn}(x)$ of curved space
or in terms of a vierbein $e^{a}_m(x)$
(where $a$ is a tangent space index) since the number of independent
components of $e^{a}_m(x)$ coincides with that of the metric due to
the local Lorentz invariance in the tangent space. However, if the
local symmetry in the tangent space is smaller than Lorentz symmetry
there is a difference in the number of the components of the
metric and of the vierbein, and gravity theory formulated in terms of the
vierbein will differ from that described by the metric.
For example, in our case of the D=4 model, where in the tangent space SO(1,3)
is broken to SO(3) the vierbein has three more independent components in
comparison with the metric. It is just these components of the vierbein
which remain when one considers the duality symmetric model
in the flat space--time, and
these correspond to the three independent components of $u_m(x)$.

Of course, all the consequences of this specific
local Lorentz symmetry breaking
must be understood yet (or, perhaps, there may exist a hidden
invariance, as one written in eqs. \p{sy}, which substitutes local
Lorentz symmetry), but
we stress that in the flat limit, due to the presence of $u_m(x)$,
Lorentz invariance in space--time does take place, and the auxiliary
fields decouple, as has been demonstrated above.

With all these points in mind we propose a general coordinate invariant
action, which describes the coupling of the duality symmetric gauge fields
and a Majorana fermion $\psi^\mu(x)$ to
gravity, in the following form:
$$
S=\int d^4x\det({e^{a}_m}) \left(-{1\over 8}F^\a_{mn}F^{\a mn}+
{1\over 4}e^m_{(o)}{\cal F}^\a_{ml}g^{ln}{\cal F}^{\a}_{np}e^p_{(o)}
-{1\over{\det{e^{a}_m}}}e^{(o)}_p\e^{plmn}\partial_l\L_{mn}+R\right)
$$
\begin{equation}\label{gra}
-i\int d^4x(\det{e^{a}_m})\bar\psi(x)\g^ae^m_a(\partial_m
+{1\over 4}\om^{bc}_m\g_a\g_b)\psi(x).
\end{equation}
where $g_{mn}\equiv e^{a}_me_{an}$ is a metric, $R(x)$ is the scalar
curvature of the curved space--time and $\om^{ab}_m$ is the $SO(1,3)$ spin
connection. When the metric is flat, $e^{a}_m$ reduce to the local
Lorentz frame fields $u^a_m=(u_m, u^i_m)$ (eqs. \p{u}, \p{or}) and
$\om^{ab}_m=u^a_n\partial_mu^{bn}$, the first
raw of eq. \p{gra} reduces to \p{gen} and the fermion part becomes
the free fermion action
\begin{equation}\label{free}
S=-i\int d^4x\overline{\hat\psi}\g^m\partial_m\hat\psi
\end{equation}
for a redefined fermion field
$\hat\psi^{\mu}(x)=v^{\mu}_{\nu}(x)\psi^\nu(x)$, where
$v^{\mu}_{\nu}(x)$ is a $Spin(1,3)$ matrix related to $ u^a_m(x)$ through the
well known $Spin(1,3)$--transformation law for the gamma--matrices:
$$
v^{-1}\g^mv=u^m_a\g^a.
$$
In agreement with a supersymmetric version of \cite{ss},
the flat limit of the action \p{gra} is supersymmetric under the
following transformations of $A^\a_m$ and $\hat\psi$:
$$
\d A^\a_m=i{\overline{\hat\psi}}\g_m\e^\a,
$$
\begin{equation}\label{susy}
\d{\hat\psi}={1\over 8}F^{\a mn}\g_m\g_n\e^\a+
{1\over 4}u_p{\cal F}^{\a
pm}u^n\g_m\g_n\e^\a,
\end{equation}
where $\e^\a$ are constant Majorana spinors subject to the
condition $\e^\a=i\g_5{\cal L}^{\a\b}\e^\b$.
On the mass shell, when ${\cal F}^\a_{mn}=0$ (eq. \p{dual}),
the transformations \p{susy} reduce to the
standard supersymmetry transformations for a vector supermultiplet
($A^1_m,\hat\psi$).

The model considered above
admits the generalization to duality symmetric models
of antisymmetric gauge fields in various dimensions \cite{ss}
and may be regarded as a basis
for constructing manifestly--covariant duality--symmetric
effective actions coupled to supergravity.

\bigskip
{\bf Acknowledgements}. D.S. is grateful to John Schwarz for
discussion on duality, and to Igor Bandos and Dmitrij Volkov for constant
interest to this work and fruitful discussion. P.P., D.S. and M.T. would
also like to thank Francesco Toppan for illuminative questions.

Work of D.S. was supported in part by  the
International Science Foundation under the grant N RY 9000,
 by the State  Committee  for  Science  and  Technology  of
Ukraine under the Grant N 2/100 and by the
INTAS grants 93--127, 93--493, 93--633.


\begin{thebibliography}{99}
\bibitem{ss}
J. H. Schwarz and A. Sen, \NPB 411 1994 35.
\bibitem{l}
A. Khoudeir and N. Pantoja, hep--th/9411235.
\bibitem{gs}
M. Green and J. H. Schwarz, \PLB 109 1982 444; \PLB 131 1984 367; \NPB
243 1984 475.
\bibitem{gikos}
A.  Galperin, E. Ivanov, S.  Kalitzin, V. Ogievetsky and E. Sokatchev,
\CQG 1 1984 498; \CQG 2 1985 155.
\end{thebibliography}
\end{document}